\documentclass{iau}

\usepackage{graphicx}

\newcommand\aj{\textit{AJ}}
\newcommand\apj{\textit{ApJ}}
\newcommand\apjs{\textit{ApJS}}
\newcommand\aap{\textit{A$\&$A}}
\newcommand\mnras{\textit{MNRAS}}
\newcommand\nat{\textit{Nature}}
\newcommand\jcap{\textit{JCAP}}

\title[Peak/Dip Picture]{The Peak/Dip Picture of the Cosmic Web}
\author[G. Rossi]{Graziano Rossi}
\affiliation{Department of Astronomy and Space Science, Sejong University, Seoul, 143-747, Korea \\ email: {\tt graziano@sejong.ac.kr}}
\pubyear{2014} 
\volume{308}  
\jname{The Zeldovich Universe: Genesis and Growth of the Cosmic Web}
\editors{Rien van de Weygaert, Sergei Shandarin, \\ Enn Saar \& Jaan Einasto, eds. \hspace{3cm}}

\begin{document}
\maketitle



\begin{abstract}
The initial shear field plays a central role in the formation of large-scale structures, and in shaping the geometry, morphology, and topology of the cosmic web.
We discuss a recent theoretical framework for the shear tensor, termed the `peak/dip picture', which accounts for the fact that halos/voids may form from local extrema of the density field -- rather than from random spatial positions;
the standard Doroshkevich's formalism is generalized, 
to include correlations between the density Hessian and shear field 
at special points in space 
around which halos/voids may form. We then present the `peak/dip excursion-set-based' algorithm, along with its most recent applications -- merging peaks theory with the standard excursion set approach.   

\end{abstract}



\firstsection
\section{The Cosmic Web: Geometry, Morphology, Topology}

\begin{figure}[t]
\begin{center}
\includegraphics[width=2.63in]{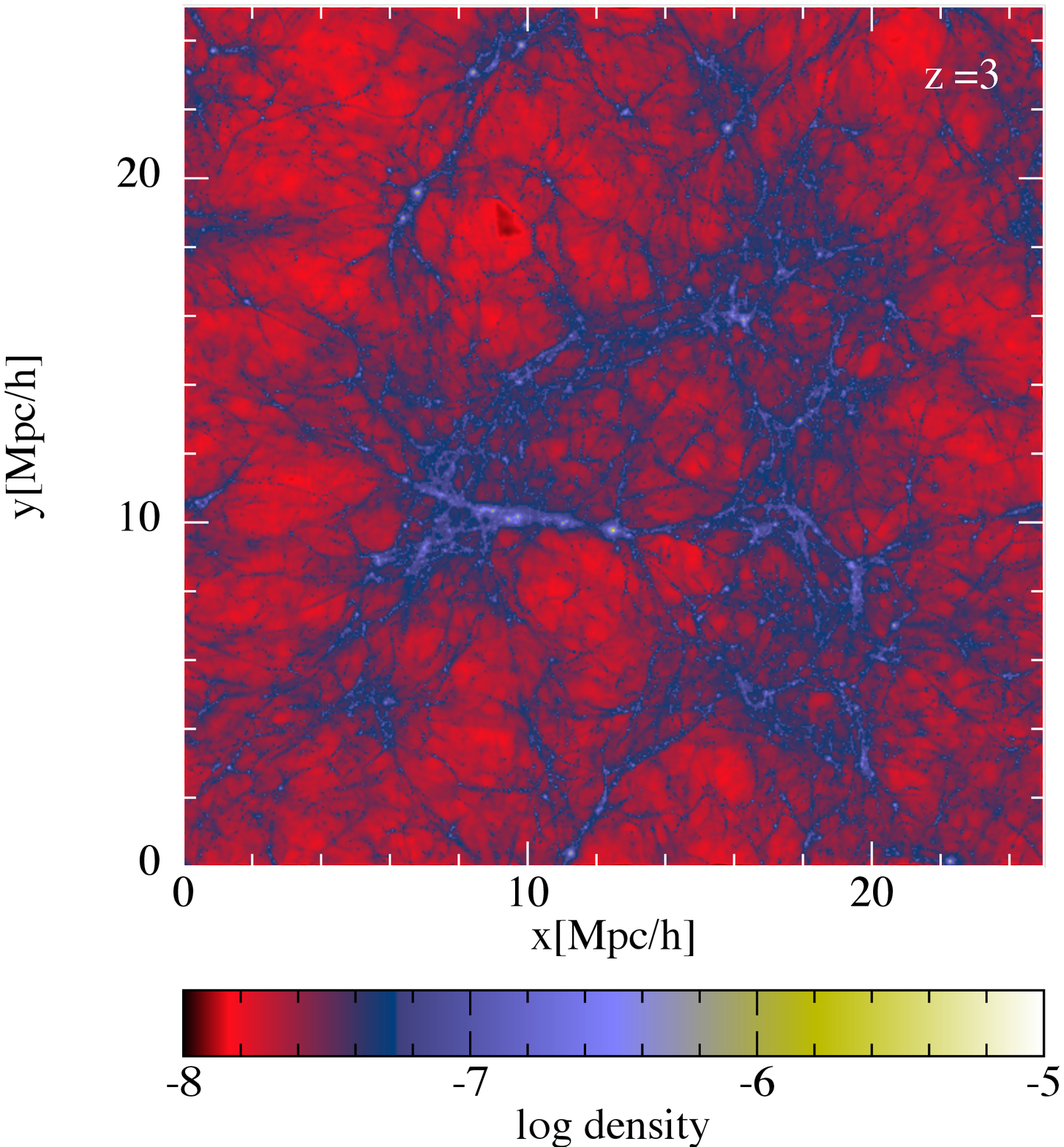} 
\includegraphics[width=2.63in]{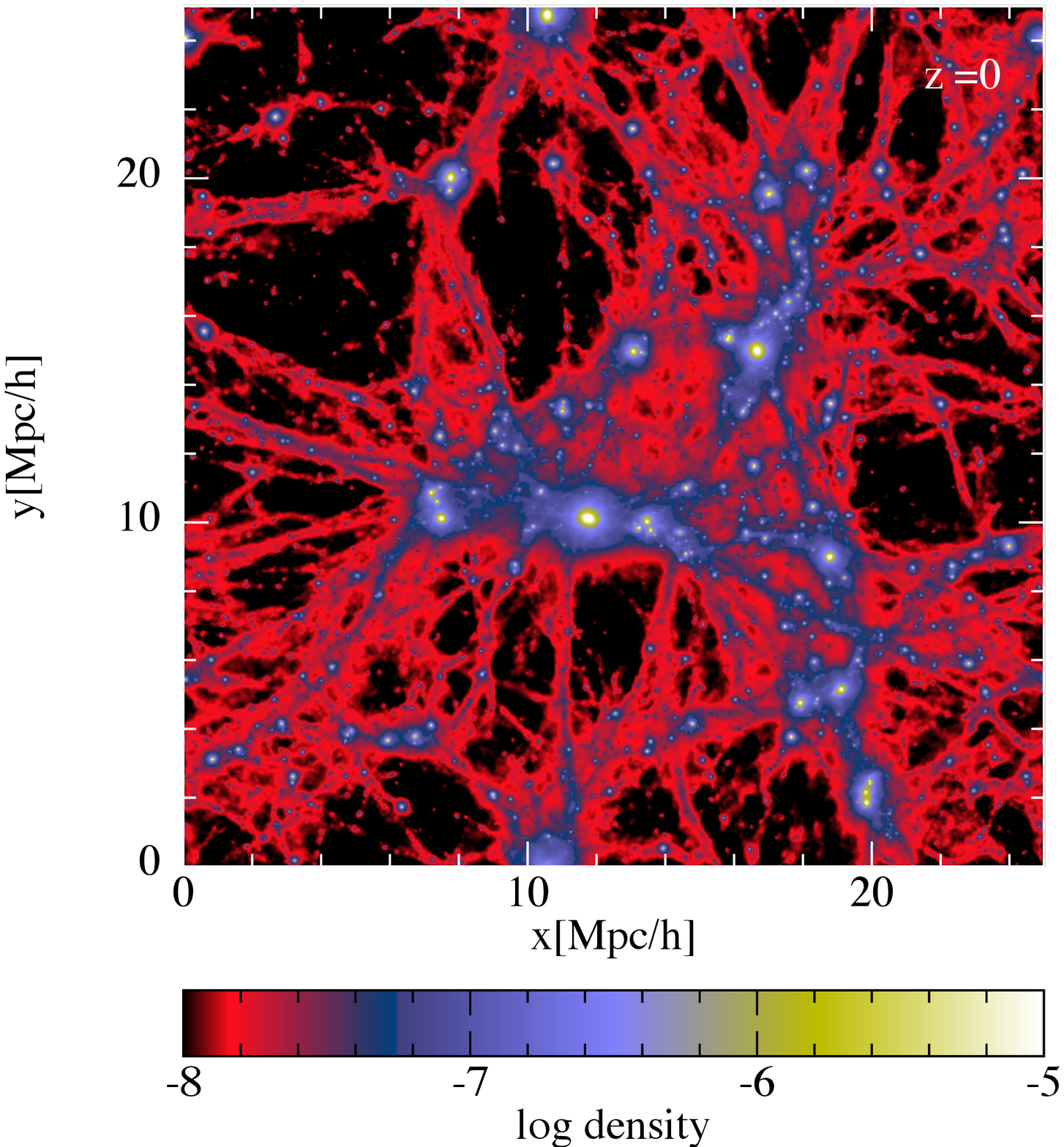} 
\includegraphics[width=2.63in]{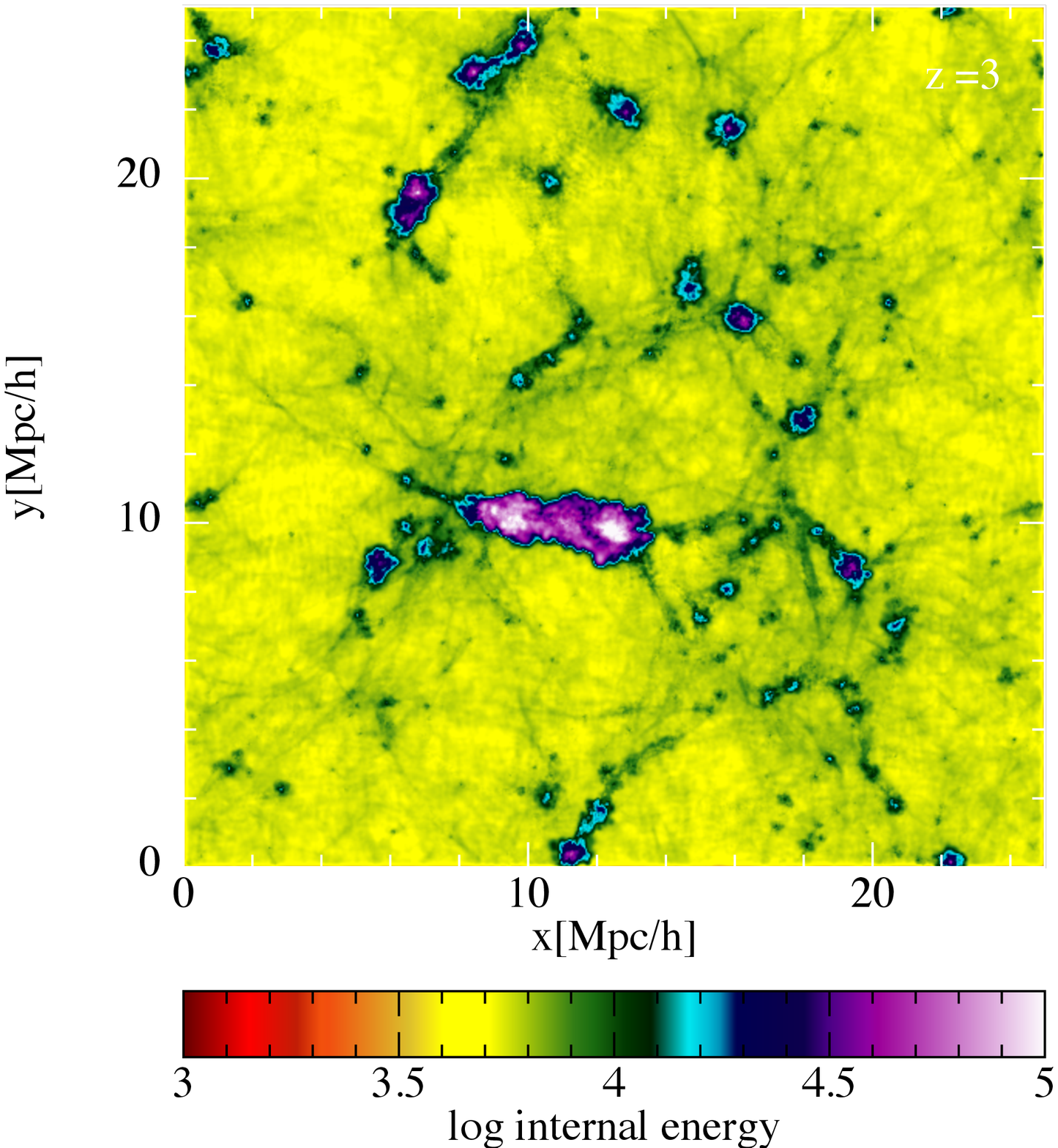} 
\includegraphics[width=2.63in]{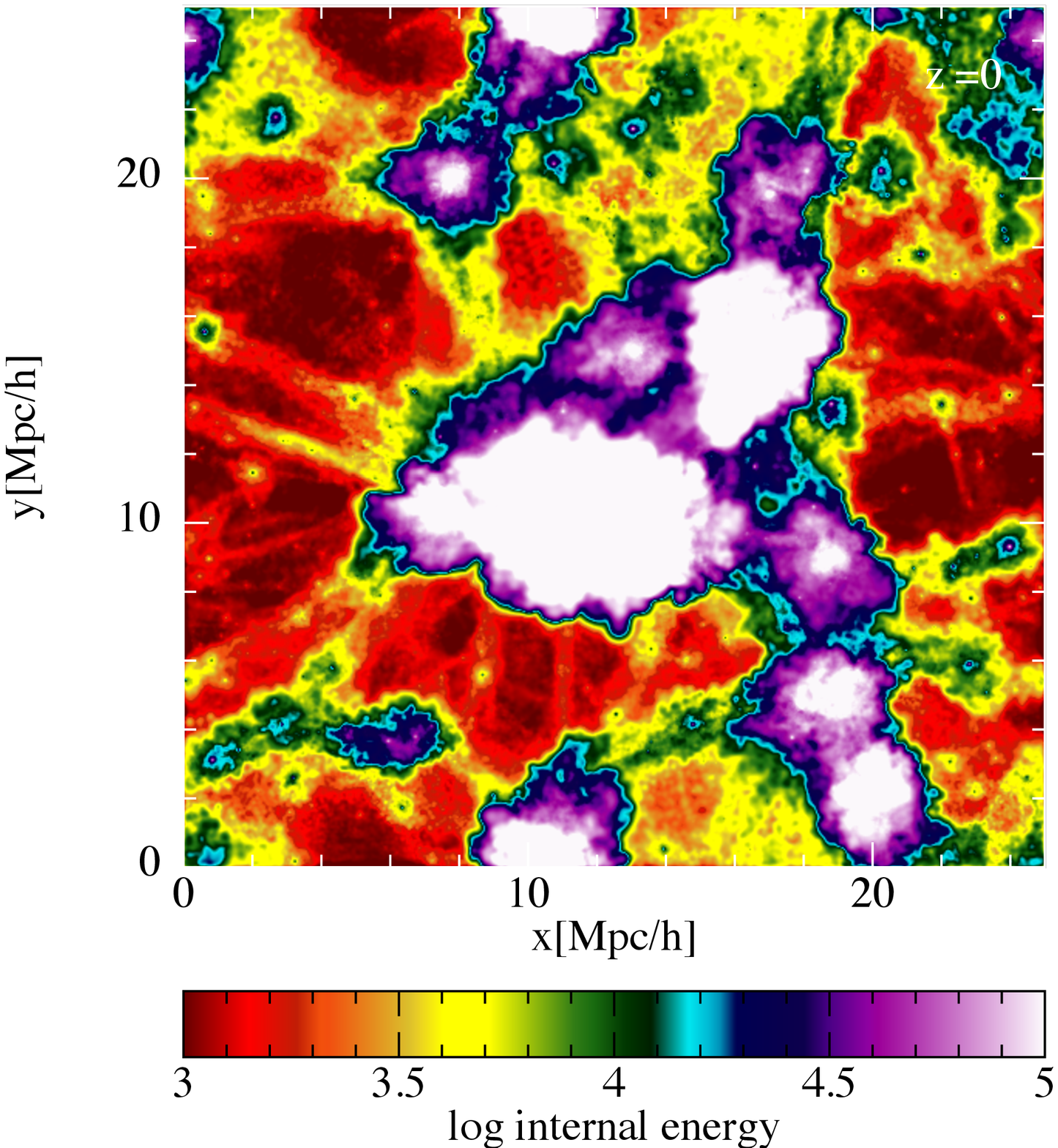} 
\caption{Snapshots of the gaseous component of the cosmic web, from a cosmological hydrodynamical simulation with $25h^{-1}$Mpc box-size 
and resolution $N_{\rm p}=192^3$ particles/type, assuming a 
reference Planck (2013) cosmology. Top panels are full projections of the density field in the $x$ and $y$ directions, at redshifts $z=3$ (left) and $z=0$ (right). Bottom panels show the corresponding 
internal energy distributions, related to the temperature of the gas.}
\label{fig1}
\end{center}
\end{figure}

The `cosmic web', a
complex large-scale spatial organization of matter, is
the result of the anisotropic nature of gravitational collapse (\cite[Zel'Dovich 1970]{1970A&A.....5...84Z}; \cite[Peebles 1980]{1980lssu.book.....P}; \cite[Bardeen et al. 1986]{1986ApJ...304...15B}; \cite[Bond et al. 1991]{1991ApJ...379..440B}). 
Figure \ref{fig1} exemplifies this intricate pattern of filaments, halos, voids and sheets, in its gaseous component at $z=3$ (top left panel) -- when structures are still forming -- and
at the present epoch ($z=0$; top right panel), when the geometry, morphology, and topology of the web are well-delineated; these snapshots are extracted from
a 25$h^{-1}$Mpc box-size  low-resolution hydrodynamical simulation, assuming a Planck (2013) reference cosmology (\cite[Borde et al. 2014]{2014JCAP...07..005B}; \cite[Rossi et al. 2014]{2014A&A...567A..79R}).
In the same figure, the bottom panels show the corresponding distribution of the internal energy at the two different redshifts considered.
The cosmic web has been confirmed by several observations, in particular using data from the 2dF Galaxy Redshift Survey (Colless et al. 2003) and the Sloan Digital Sky Survey (York et al. 2000), and most recently
by a spectacular three-dimensional detection of a cosmic web filament in Lyman-$\alpha$ emission at $z\simeq 2.3$, discovered during a search for  cosmic gas fluorescently illuminated by background bright quasars (Cantalupo et al. 2014).
Along with observational efforts, numerical studies are essential and unavoidable in order to understand
all the complicated physical phenomena involved in the nonlinear formation of structures, requiring larger simulated volumes and yet accuracy in resolution 
 -- e.g., see \cite[Cautun et al. (2014)]{2014MNRAS.441.2923C} and \cite[Hidding et al. (2014)]{2014MNRAS.437.3442H} for recent developments.
Some analytic work is also helpful in guiding and interpreting results from numerical studies.
To this end, we discuss here a theoretical framework which patches together several ingredients: the fact that dark matter halos and voids are clearly not spherical in shape, but at the very least triaxial; the observation, supported by simulation results,  
that  halos and voids tend to form in or around local maxima/minima of the density field -- rather than at random spatial locations; the indication that
an `embryonic cosmic web' is  already present in the primordial density field (Bond et al. 1991);
the evidence that there is good correspondence between 
peaks/dips in the initial conditions and halos/voids at later times; the crucial role of the tidal field in shaping the cosmic web, and the effects of its correlation with the density Hessian; 
and the fact that the standard excursion set theory, which determines the initial conditions and is based on the statistics of Gaussian fields, 
only considers random spatial positions. The aim is to provide a more realistic theoretical language for describing the morphology of the cosmic web, in closer support of numerical studies. We briefly elaborate on these points in the next sections,
while details can be found in \cite[Rossi et al. (2011)]{2011MNRAS.416..248R} and in \cite[Rossi (2012, 2013)]{2012MNRAS.421..296R-2013MNRAS.430.1486R}.


\section{Initial Shear Field and Dynamical Evolution}

\begin{figure}[t]
\begin{center}
\includegraphics[width=5.6in]{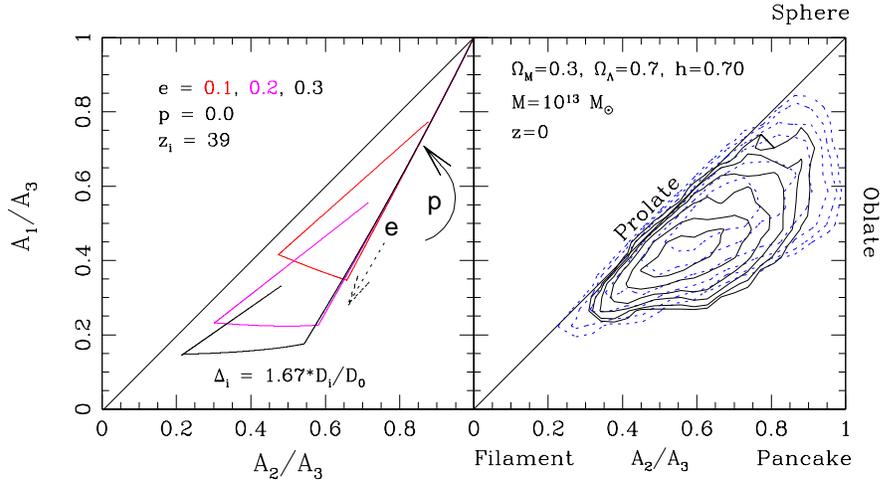} 
\caption{A simple model for describing the shapes of dark matter halos (\cite[Rossi et al. 2011]{2011MNRAS.416..248R}), based on the combination of
excursion set theory and ellipsoidal collapse. The left panel shows the evolution of halos in the `axis ratio plane', as a function of their
ellipticity $e$ and prolateness $p$, when $p=0$ and $e=0.1, 0.2, 0.3$, respectively. 
The right panel compares the final axial ratio distributions predicted
from the model (black solid lines) with corresponding results from numerical simulations (blue dashed lines), for 10,000 halos of mass  $M= 10^{13} M_{\rm \odot}$.}
\label{fig2}
\end{center}
\end{figure}

The initial shear field -- rather than the density Hessian --
is a major player in shaping the geometrical structure of the cosmic web, and originates
the characteristic observed pattern of  filaments, halos, voids, and sheets.
The morphology of halos and voids departs from sphericity, as the statistics of Gaussian fields
imply that spherically symmetric initial configurations should be a set of measure zero (\cite[Doroshkevich 1970]{1970Afz.....6..581D}); the observed triaxiality has
several important implications in determining the assembly histories, kinematics, clustering and fundamental structural
properties of halos and voids. 
Rossi et al. (2011) described a simple dynamical model able to link the final shapes of virialized halos to their initial shapes, by combining the
physics of the ellipsoidal collapse with the excursion set theory. 
An illustration is provided in Figure \ref{fig2}: in a useful planar representation of halo axial ratios, the left panel shows the effect of 
the initial values of ellipticity  and prolateness ($e, p$) -- which parametrize the surrounding shear field -- in determining the future evolution of an object of given mass, for three representative values of $e$ when $p=0$; 
the right panel compares the distribution of final axial ratios in the model (solid black lines) with results from numerical simulations 
(dashed blue lines) when 10,000 halos of mass $M= 10^{13} M_{\rm \odot}$ are considered -- in a flat cosmology with $\Omega_{\rm M}=0.3$ and $h=0.7$.


\section{The Peak/Dip Formalism: Theory and Applications}

\begin{figure}[t]
\begin{center}
\includegraphics[width=5.6in]{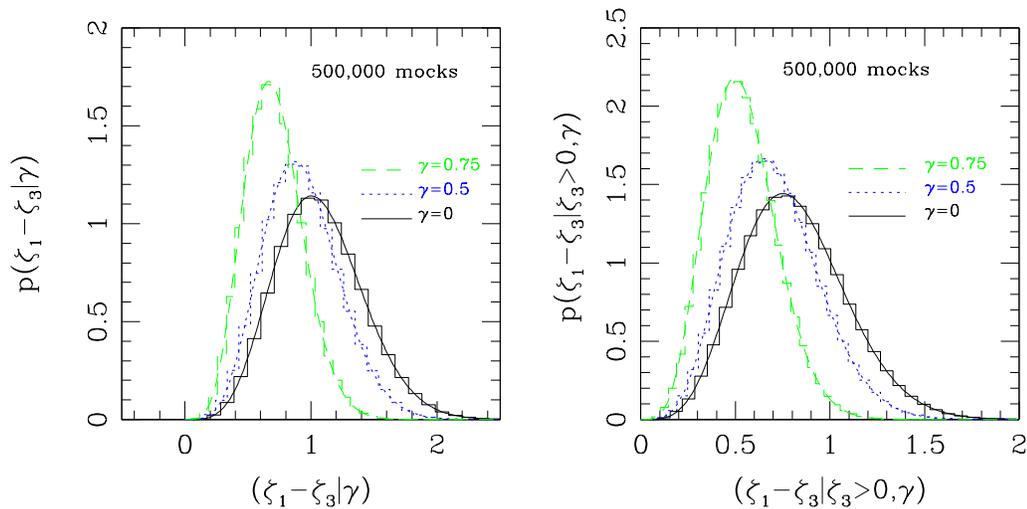} 
\caption{Conditional distributions of eigenvalues in the `peak/dip picture' relevant for the conditional shape distributions (Rossi 2012, 2013), for different values of the
correlation parameter $\gamma$ -- as specified in the panels. Theoretical predictions are confronted with results from 500,000 mock realizations (histograms), obtained with the
`peak-dip excursion-set-based algorithm'.}
\label{fig3}
\end{center}
\end{figure}

Although idealized, the previous model provides some useful insights into the nonlinear collapse of structures, and in particular it highlights the fact that 
a collapsing patch will eventually become a filament, pancake, or halo depending on its
 initial shape and overdensity. Hence, the 
eigenvalues of the initial shear field are a key ingredient in determining the final destiny of an object, and 
more generally in shaping the morphology of the cosmic web. 
In 1970, Doroshkevich derived 
 the joint `unconditional' probability distribution of an ordered set of tidal field eigenvalues corresponding to a Gaussian potential. 
Recently, akin in philosophy to that of
van de Weygaert \& Bertschinger (1996), Rossi (2012) provided a set of analytic expressions which extended
the work of Doroshkevich (1970) and Bardeen et al. (1986),
to incorporate the density
peak/dip constraint into the statistical description of the 
initial shear field. In this generalized formalism, termed the `peak/dip picture of the cosmic web', the probability of observing a tidal field ${\bf T}$ for the gravitational potential given a curvature
${\bf H}$ for the density field and a correlation strength $\gamma$ is given by (Rossi 2012):
\begin{equation}
p ({\bf T}|{\bf H},\gamma) = {15^3 \over 16 \sqrt {5} \pi^3} {1 \over \sigma^6_{\rm T}(1-\gamma^2)^3} {\rm exp} \Big [-{3 \over 2 \sigma_{\rm T}^2(1-\gamma^2)} (2 K_1^2 - 5 K_2 ) \Big ].
\label{doro_inter_extended}
\end{equation}
The previous equation has lead to the `peak/dip excursion-set-based' algorithm,
able to sample the constrained eigenvalues of the initial shear field associated with Gaussian statistics  
at positions which correspond to peaks or dips of the correlated density field (Rossi 2013); the algorithm can be
 readily inserted into the excursion set framework (Peacock \& Heavens 1990; Bond et al. 1991; Lacey \& Cole 1993) 
to account for a subset of spatial points from where halos/voids may form -- hence merging the peaks theory description with the excursion set approach. 
Along these lines, it is possible to
extend the standard distributions of shape parameters (i.e. ellipticity and prolateness) in the presence of the density peak/dip constraint (Rossi 2013, and Figure \ref{fig3}), 
which generalize some previous literature work and combine
the formalism of Bardeen et al. (1986) -- based on the density field  -- with that of Bond \& Myers (1996) -- based on the shear field.


\section*{Acknowledgments}

This work and the participation to the IAU Symposium 308  `The Zeldovich Universe: Genesis and Growth of the Cosmic Web' (June 2014) in Tallinn, Estonia, were supported by the faculty research fund of Sejong University in 2014, and by 
the National Research Foundation of Korea through SGER Grant 2014055950.
It is a pleasure to thank Sergei Shandarin and Rien van de Weygaert for the superb organization, along with 
the scientific and local organizing committees and the secretariat. 




\end{document}